\documentclass[preprint,showpacs,preprintnumbers,amsmath,amssymb]{revtex4}

\usepackage{graphicx}
\usepackage{dcolumn}
\usepackage{bm}

\begin{document}

\title{Detection of source inhomogeneity through event-by-event two-pion
Bose-Einstein correlations}

\author{Yan-Yu Ren$^1$}
\author{Wei-Ning Zhang$^{1,2}$\footnote{wnzhang@dlut.edu.cn}}
\author{Jian-Li Liu$^{1}$}

\affiliation{
$^1$Department of Physics, Harbin Institute of Technology,
Harbin, Heilongjiang 150006, China\\
$^2$School of Physics and Optoelectronic Technology, Dalian University
of Technology, Dalian, Liaoning 116024, China\\
}

\date{\today}

\begin{abstract}
We develop a method for detecting the inhomogeneity of the
pion-emitting sources produced in ultra-relativistic heavy ion
collisions, through event-by-event two-pion Bose-Einstein
correlations.  The root-mean-square of the error-inverse-weighted
fluctuations between the two-pion correlation functions of single
and mixed events are useful observables for the detection.  By
investigating the root-mean-square of the weighted fluctuations for
different impact parameter regions people may hopefully determine
the inhomogeneity of the particle-emitting in the coming Large
Hadron Collider (LHC) heavy ion experiments.
\end{abstract}


\maketitle

Two-pion Hanbury-Brown-Twiss (HBT) interferometry is a useful tool
for probing the space-time structure of the particle-emitting
sources produced in high energy heavy ion collisions
\cite{Won94,Wie99,Wei00,Lis05}.  Because of the limitation of data
statistics, the usual HBT investigations are performed for mixed
events and the HBT radii are obtained by fitting the two-pion HBT
correlation functions from the mixed events to Gaussian parametrized
formulas.

On event-by-event basis, the density distribution of the source may
be not a Gaussian distribution.  An inhomogeneous particle-emitting
source on event-by-event basis may be a more general case because of
the fluctuating initial matter distribution in high energy heavy ion
collisions \cite{Gyu97,Dre02,Ham04,And08}.  In Ref. \cite{Zha06}, a
granular source model was used to explain the Relativistic Heavy Ion
Collider (RHIC) HBT results, $R_{\rm out} / R_{\rm side} \approx 1$
\cite{STA01a,PHE02a,PHE04a,STA05a}. The main idea of the explanation
is that the evolution time for the granular sources is short and the
short evolution time, which can not be averaged out after
event-mixing, may lead to the HBT results $R_{\rm out} \sim R_{\rm
side}$ \cite{Zha06,Zha04,Zha07}.  Recent source imaging researches
for the collisions of $\sqrt{s_{_{\rm NN}}}=200$ GeV Au+Au indicate
that the pion-emitting source with the selections $40 < {\rm
centrality} < 90\%$ and $0.20 < k_{\rm T} < 0.36$ GeV/c is far from
a Gaussian distribution \cite{PHE07,Bro07,Yan08}.  Although the long
tail of the two-pion source function at large separation $r$ is
believed mainly the contribution of long-lived resonances
\cite{PHE07,Bro07}, the enhancement of the source function at small
$r$ may possibly arise from the source inhomogeneity \cite{Yan08}.

For inhomogeneous particle-emitting sources, the single-event
two-pion HBT correlation functions may exhibit fluctuations relative
to the HBT correlation functions of mixed events \cite{Won04,Zha05}.
Detecting and investigating this event-by-event fluctuations is
important for finally determining the source inhomogeneity and
understanding the initial conditions and evolution of the system in
high energy heavy ion collisions.

Hydrodynamics may provide a direct link between the early state of
the system and final observables and has been extensively used in
high energy heavy ion collisions.  In hydrodynamical calculations
the system evolution is determined by the initial conditions and the
equation of state (EOS) of the system.  Smoothed Particle
Hydrodynamics (SPH) is a suitable candidate that can be used to
treat the system evolution with large fluctuating initial conditions
for investigating event-by-event attributes \cite{Ham04,Agu01}. It
has been used in high energy heavy ion collisions for a wide range
of problems \cite{Ham04,Agu01,Gaz03,Soc04,Ham05,And06,And08}. In the
present letter we use SPH to describe the system evolution. The
system initial states are given by the NEXUS event generator
\cite{Dre02} at $\tau_0=1$ fm/c for $\sqrt{s_{_{\rm NN}}}=200$ GeV
Au+Au collisions at RHIC and $\tau_0=0.5$ fm/c for $\sqrt{s_{_{\rm
NN}}}=5500$ GeV Pb+Pb collisions at LHC.  The EOS is obtained with
the entropy density suggested by QCD lattice results
\cite{Bla87,Lae96,Ris96,Zha04}.  In the EOS, the QGP phase is
considered as an ideal gas of massless quarks (u, d, s) and gluons
\cite{Gaz03,Soc04}.  The hadronic gas is composed of the resonances
with mass below 2.5 GeV/c$^2$, where volume correction is taken into
account \cite{Gaz03,Soc04}.  The transition temperature between the
QGP and hadronic phases is taken to be $T_{\rm c}=160$ MeV, and the
width of the transition is taken to be $0.1T_{\rm c}$ \cite{Ris96}.

The coordinates used in SPH are $\tau=\sqrt{t^2-z^2},\,x,\,y,$ and
$\eta=(1/2)\ln [(t+z)/(t-z)]$ \cite{Ham04,Agu01,Gaz03}.  They are
convenient for describing the system with rapid longitudinal
expansion. However, in order to investigate the whole space-time
structure of the system an nonlocal coordinate frame is needed and
we work in the center-of-mass frame of the system. Figure 1(a), (b),
and (c) show the pictures of the transverse distributions of energy
density for one $\sqrt{s_{_{\rm NN}}}=200$ GeV Au+Au event at $t=1$
fm/c and with impact parameter $b=0$, 5, and 10 fm, respectively.
The pictures are taken for the spatial region ($|x,y|<12$ fm,
$|z|<0.83$ fm) and with an exposure of $\Delta t=0.3$ fm$/c$.  Fig.
1(a$'$), (b$'$), and (c$'$) are the corresponding pictures taken at
$t=5$ fm/c with the same $\Delta t$ and for the spatial region
($|x,y|<12$ fm, $|z|<1$ fm).  One can see that the systems are
inhomogeneous in space and time.  There are many ``lumps" in the
systems and the number of the lumps decreases with impact parameter
increasing.

\begin{figure}
\vspace*{-5mm}
\includegraphics[angle=0,scale=0.33]{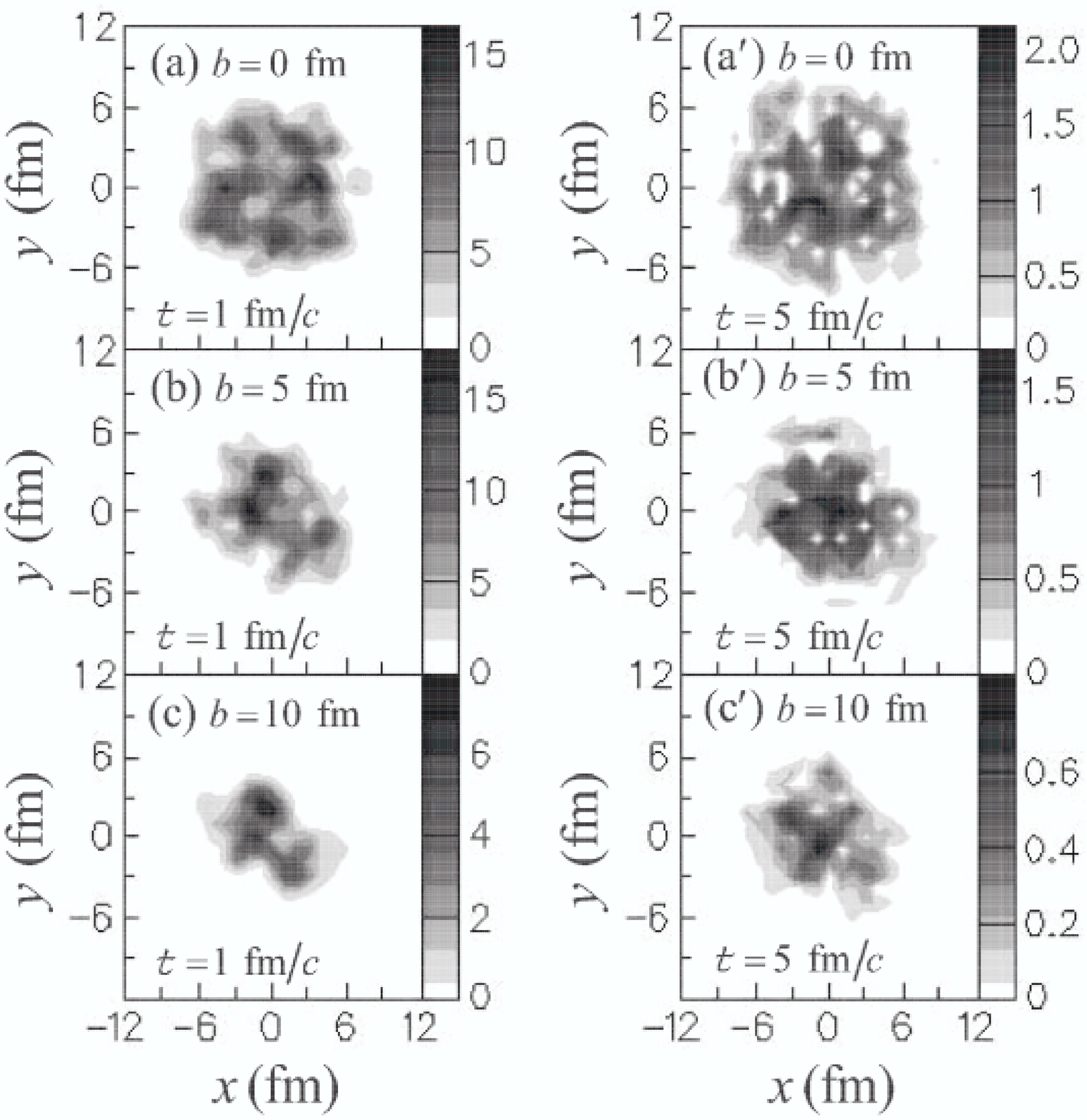}
\vspace*{-15mm} \caption{\label{fig:fig1} Transverse distributions
of energy density of the systems with different impact parameters
and at $t=1$ fm/c [(a)--(c)] and $t=5$ fm/c [(a$'$)--(c$'$)]. The
unit of energy density is GeV/fm$^3$.}
\end{figure}

Assuming that final identical pions are emitted at the space-time
configuration characterized by a freeze-out temperature $T_{\rm f}$,
we may generate the pion momenta according to Bose-Einstein
distribution and construct the single-event and mixed-event two-pion
correlation functions \cite{Zha06,Zha05}.  Figure 2 shows the
two-pion correlation functions $C(q_{\rm side}, q_{\rm out}, q_{\rm
long})$ for the single and mixed events for $\sqrt{s_{_{\rm
NN}}}=200$ GeV Au+Au with impact parameters $b=$ 10 fm (up panels)
and $b=$5 fm (down panels). Here $q_{\rm side}$, $q_{\rm out}$, and
$q_{\rm long}$ are the components of ``side", ``out", and ``long" of
relative momentum of pion pair \cite{Pra86,Ber88}.  In each panel of
Fig. 2, the dashed lines give the correlation functions for a sample
of different single events and the solid line is for the mixed event
obtained by averaging over 40 single events. In our calculations,
the freeze-out temperature is taken to be 150 MeV. For each single
event the total number of generated pion pairs in the relative
momentum region $(q_{\rm side}, q_{\rm out}, q_{long} \leq 200$
MeV/c) is $N_{\pi\pi}=10^6$ and the numbers of the pion pairs in the
relative momentum regions $(q_i \leq 200\, {\rm MeV/c};\, q_j,\, q_k
\leq 30\, {\rm MeV/c})$ are about $2.7\%N_{\pi\pi}$, where $i$, $j$,
and $k$ denote ``side", ``out", and ``long".

\begin{figure}
\includegraphics[angle=0,scale=0.35]{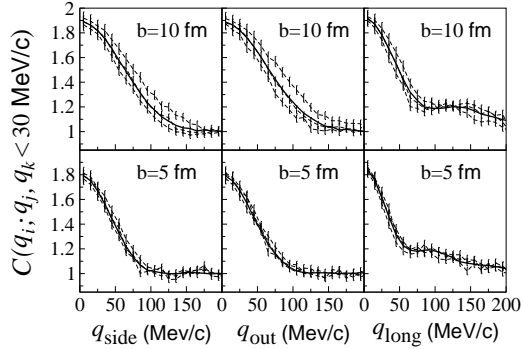}
\caption{\label{fig:fig2} Two-pion correlation functions for a
sample of different single events (dashed lines) and mixed events
(solid lines) with impact parameters $b=$10 fm (up panels) and $b=$5
fm (down panels).}
\end{figure}

From Fig. 2 it can be seen that the correlation functions for the
single events exhibit fluctuations relative to those for the mixed
events. These fluctuations are larger for bigger impact parameter.
It is because that the number of the lumps in the system decreases
with impact parameter and the fluctuations are larger for the source
with smaller number of lumps \cite{Won04,Zha05}.  It also can be
seen that in the longitudinal direction the correlation functions
exhibit oscillations which can not be smoothed out by event mixing.
It is because that there are two sub-sources moving forward and
backward the beam direction.  This oscillations of the mixed-event
correlation functions will not appear if we apply an additional cut
for the initial rapidity of the ``smoothed particles", $\eta_0
>0$ or $\eta_0 < 0$.

We have seen that the two-pion correlation functions of the single
events exhibit event-by-event fluctuations.  However, in the usual
mixed-event HBT measurements, these fluctuations are smoothed out.
In order to observe the event-by-event fluctuations, we investigate
the distribution $dN/df$ of the fluctuations between the correlation
functions of single and mixed events, $|C_s(q_i) - C_m(q_i)|$, with
their error-inverses as weights \cite{Zha05},
\begin{equation}
\label{RF} f(q_i)  = \frac{|C_s(q_i) - C_m(q_i)|}{\Delta |C_s(q_i) -
C_m(q_i)|} \,.
\end{equation}
In calculations we take the width of the relative momentum $q_i$ bin
to be 10 MeV and use the bins in the region $20 \le q_i \le 200$
MeV. The up panels of Fig. 3 show the distributions of $f$ in the
``side", ``out", and ``long" directions, obtained from 40 simulated
$\sqrt{s_{_{\rm NN}}}=200$ GeV Au+Au events. The impact parameter
for these events is $b=5$ fm and the number of correlated pion pairs
for each of these events is $N_{\pi\pi}=10^7$.  The solid lines are
the results for the events with the fluctuating initial conditions
(FIC) generated by NEXUS. Because in the last analysis the
fluctuations of the single-event correlation functions are from the
FIC, for comparison we also investigate the distribution $dN/df$ for
the events with the smooth initial conditions (SIC) obtained by
averaging over 30 random NEXUS events \cite{Ham04,Soc04,Ham05,And08}
with the same $b$ as for the FIC. It can be seen that the $f$
distributions for FIC are much wider than the corresponding results
for SIC.  The down panels of Fig. 3 show the distributions of $f$
for the 40 simulated events and $N_{\pi\pi}=5\times10^6$. It can bee
seen that the widths of the distributions for FIC decrease with
$N_{\pi\pi}$ decreasing and for $N_{\pi\pi}=5\times10^6$ the
distributions for FIC are still wider than those for SIC.

\begin{figure}
\includegraphics[angle=0,scale=0.3]{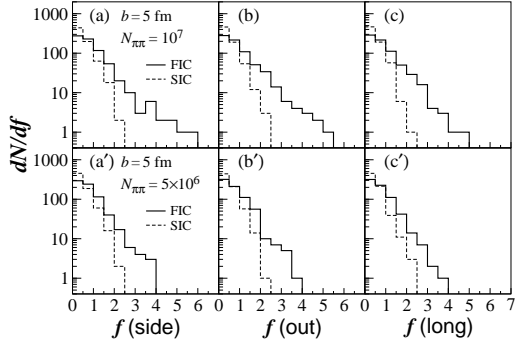}
\caption{\label{fig:fig3} The distributions $dN/f$ for 40 events
with FIC and SIC.  $b=5$ fm.}
\end{figure}

\begin{figure}
\vspace*{3mm}
\includegraphics[angle=0,scale=0.42]{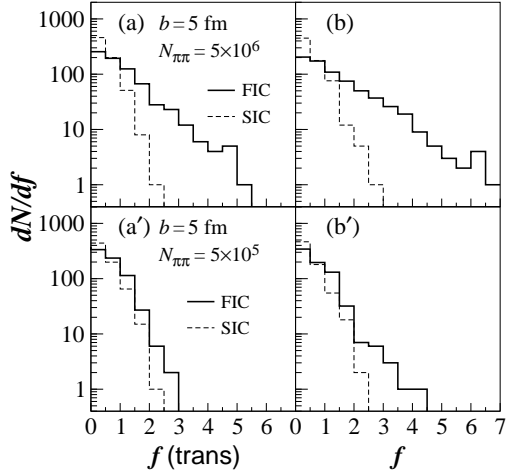}
\vspace*{3mm} \caption{\label{fig:fig4} The distributions $dN/f$ for
40 events with FIC and SIC, where $f$ are calculated for variables
$q_{\rm trans}$ and $q$.}
\end{figure}

In experiments the number of correlated pion pairs in one event,
$N_{\pi\pi}$, is limited.  It is related to the energy
$\sqrt{s_{_{\rm NN}}}$ of the collisions.  For a finite $N_{\pi\pi}$
we have to reduce variable numbers in analysis although it will lose
some details.  In Fig. 4 we show the distributions of $f$ for the
variables of transverse relative momentum $q_{\rm trans}$ and
relative momentum $q$ of the pion pairs for the 40 simulated events
with $b=5$ fm.  One can see that for $N_{\pi\pi}=5\times10^6$, the
distributions for FIC are much wider than those for SIC both for
$q_{\rm trans}$ and $q$.  Even for $N_{\pi\pi}=5\times10^5$, the
widths for FIC are visibly larger than those for SIC.  In order to
examine the distributions quantitatively, we calculate the
root-mean-square (RMS) of $f$. Figure 5 shows the RMS, $f_{\rm
rms}$, as a function of $N_{\pi\pi}$ for the 40 simulated events for
$\sqrt{s_{_{\rm NN}}}=200$ GeV Au+Au with $b=9$ fm (the up panels)
and $b=5$ fm (the down panels), where the SIC are obtained by
averaging over 100 NEXUS events.  It can been seen that the values
of $f_{\rm rms}$ rapidly increase with $N_{\pi\pi}$ for FIC because
the errors in Eq. (\ref{RF}) decrease with $N_{\pi\pi}$. For FIC the
results for $b=9$ fm are larger than the corresponding results for
$b=5$ fm because the differences $|C_s(q_i) - C_m(q_i)|$ in Eq.
(\ref{RF}) increase with $b$.  For SIC the values of $f_{\rm rms}$
are almost independent from $N_{\pi\pi}$.  It is because that both
the differences and their errors in Eq. (\ref{RF}) decrease with
$N_{\pi\pi}$ in the SIC case.

\begin{figure}
\includegraphics[angle=0,scale=0.65]{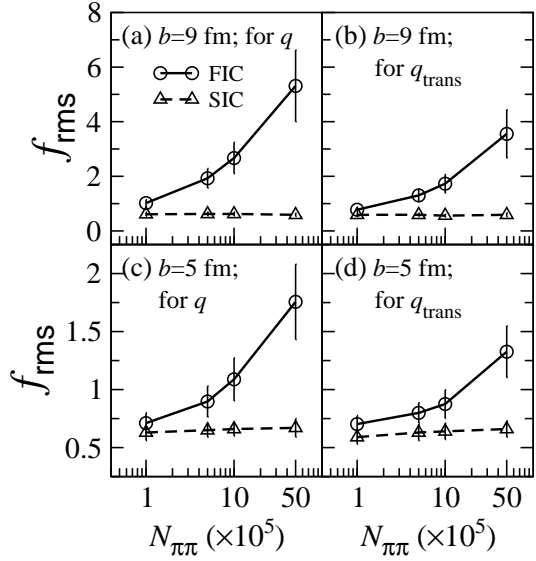}
\caption{\label{fig:fig5} The root-mean-square of $f$ for variables
$q$ and $q_{\rm trans}$ as a function of $N_{\pi\pi}$.}
\end{figure}

In Fig. 5 we only display the statistic errors which may decrease
with the number of events.  Moreover, in order to display the
fluctuations of the single-event HBT correlation functions and the
effect of FIC on the distributions of $f$, we used very large
$N_{\pi\pi}$ in our calculations.  At RHIC energy the event
multiplicity of identical pions, $M_{\pi}$, is about several
hundreds for central collisions.  The order of $N_{\pi\pi}$ is about
$10^5$ ($\sim M_{\pi}^2/2$).  However, at the higher energy of LHC,
$M_{\pi}$ will be about two thousands and the order of $N_{\pi\pi}$
will be $10^6$.  In this case the large differences between the RMS
of $f$ for inhomogeneous and homogeneous sources provide a great
opportunity to detect the source inhomogeneity.

\begin{figure}
\includegraphics[angle=0,scale=0.62]{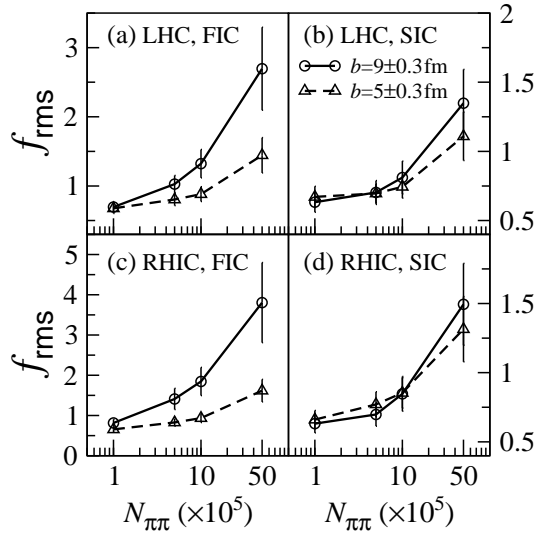}
\caption{\label{fig:fig6} The root-mean-square of $f$ calculated
with $q_{\rm trans}$ for 40 simulated events at LHC and RHIC
energies.}
\end{figure}

Another problem in experimental data analysis is that the impact
parameter $b$ is hardly to be held at a fixed value and usually be
limited in a region.  In such a case the large $f_{\rm rms}$ values
may arise from the source inhomogeneity as well as the variation of
$b$ in the region.  So one should also consider the effect of the
$b$ variation in inhomogeneity detections.

Figure 6(a) and (b) show the RMS of $f$ calculated with $q_{\rm
trans}$ for the 40 simulated events for $\sqrt{s_{_{\rm NN}}}=5500$
GeV Pb+Pb with FIC and SIC, respectively.  The impact parameters for
the FIC events are selected randomly from the regions $b=9.0\pm0.3$
(circle symbol) and $b=5.0\pm0.3$ (triangle symbol), respectively.
Whereas for each SIC event the SIC is constructed by averaging over
30 NEXUS events with the same impact parameter selected randomly
from the $b$ region as for the FIC case.  Figure 6(c) and (d) give
the corresponding results for $\sqrt{s_{_{\rm NN}}}=200$ GeV Au+Au
collisions.  Our calculations indicate that the lump numbers in the
sources at the LHC energy are larger than those at the RHIC energy.
So the RMS values for the LHC energy are smaller than those for the
RHIC energy.  From Fig. 6 one can see that for FIC case
(inhomogeneous sources) the RMS for the larger $b$ are higher than
those for the smaller $b$.  It is because that the number of lumps
in the inhomogeneous source decreases with $b$ (see Fig. 1).
However, for the SIC case the RMS for the two impact parameter
regions are almost the same for fixed $N_{\pi\pi}$.  In this case
the increases of the RMS values with $N_{\pi\pi}$ are due to the
fluctuations of the correlation functions arising from the variation
of $b$ in their regions $\delta b$.  Our calculations indicate that
these increases will become flat when $\delta b \to 0$.  So, one may
determine the source inhomogeneity by analyzing the RMS values for
different $b$ regions.  For inhomogeneous sources their differences
are bigger for larger $N_{\pi\pi}$.  Otherwise, they are almost the
same.

In summary, on event-by-event basis the initial density distribution
of matter in high energy heavy ion collisions is highly fluctuating.
The fluctuating initial conditions lead to event-by-event
inhomogeneous particle-emitting sources.  In this letter we
developed a method for detecting the source inhomogeneity through
event-by-event two-pion correlations.  We find that the RMS of the
error-inverse-weighted fluctuations $f$ are useful observables for
detecting the inhomogeneity of the sources.  The high identical pion
multiplicity in the coming LHC heavy ion collisions provides a great
opportunity to do the detections.  By investigating the RMS values
of $f$ for different impact parameter regions people may hopefully
determine the inhomogeneity of the particle-emitting produced in the
coming LHC experiments.

We thank Dr. C. Y. Wong and Dr. D. C. Zhou for helpful discussion.
This research was supported by the National Natural Science
Foundation of China under Contracts No. 10575024 and No. 10775024.

\end{document}